\documentclass[pre,aps,showpacs,superscriptaddress]{revtex4}

\usepackage{amsmath}
\usepackage{epsfig}
\usepackage[dvips]{color}

\begin{document}

\title{Splitting rate matrix as a definition of time reversal in master equation systems}
\author{Fei Liu}
\address{State Key Laboratory of Software Development Environment }
\address{School of Physics and Nuclear Energy Engineering, Beihang University, Beijing 100191, China}
\author{Hong Lei}
\address{Foreign Languages Department, Graduate University of Chinese Academy of Sciences,
China}\email{feiliu@buaa.edu.cn}

\begin{abstract}
Motivated by recent progresses in nonequilibrium Fluctuation
Relations, we present a generalized time reversal for stochastic
master equation systems with discrete states that is defined as a
splitting of the rate matrix into irreversible and reversible
parts. An immediate advantage of this definition is that a variety
of fluctuation relations can be attributed to different matrix
splitting. Additionally, we also find that, the accustomed total
entropy production formula and conditions of the detailed balance
must be modified appropriately to account for the presence of the
reversible part, which was completely ignored in the past a long
time.
\end{abstract}

\maketitle

\section{Introduction}
Fluctuation theorems or fluctuation
relations~\cite{Bochkov77,Evans,Searles,Gallavotti,Kurchan,Lebowitz,JarzynskiPRL97,
JarzynskiPRE97,Crooks99,Crooks00,HatanoSasa,Maes,SeifertPRL05,Speck}
are a variety of exact equalities about statistics of entropy
production or dissipated work that are held even in far from
equilibrium regimes. In near-equilibrium region, these relations
reduce to the famous fluctuation-dissipation theorems
(FDTs)~\cite{Evans,Lebowitz,Callen,Kubo,GallavottiPRL96}. The
discovery of these fluctuation relations significantly advances
our understanding about nonequilibrium physics, and particularly
about the second law of thermodynamics of small
systems~\cite{Bustamante}.

The fluctuation relations are very relevant with the concept of
time
reversal~\cite{Evans,Searles,Gallavotti,Kurchan,Lebowitz,JarzynskiPRL97,
JarzynskiPRE97,Crooks99,Crooks00,HatanoSasa,Maes,SeifertPRL05,Chernyak,TCohen}.
For instance, under the framework of Markovian stochastic systems,
previous work has proved that a majority of them can be derived by
a ratio of the probability densities of observing a trajectory in
a original system and the reversed trajectory in the time-reversed
system. Very recently, Chetrite and Gawedzki~\cite{Chetrite}
further elaborated this observation and presented a generalized
time reversal definition on continuous diffusion processes.
Different from conventional definition of time reversal as simply
changing time parameter in a stochastic dynamics into minus, they
explicitly defined time reversal as a splitting of drift vector
into irreversible and reversible parts which possess distinct
rules under the transformation of $t$$\to$$-t$. Because of freedom
of the splitting, a variety of time reversal and corresponding
fluctuation relations are obtained, e.g., the Hatano-Sasa
equality~\cite{HatanoSasa} arising from a novel time reversal with
nonzero reversible drift. The importance of the generalized time
reversal was shown again when we understood the origin of a
generalized integral fluctuation relation (GIFR) on general
diffusion processes~\cite{LiuF1,LiuF3}.

In addition to the continuous diffusion process, another typical
and important Markovian process is master equation with discrete
states and continuous time~\cite{Gardiner}. Although the latter is
more general than the former in principle, due to their highly
formal analogy, many results and evaluations about the fluctuation
relations in the diffusion processes could be established
correspondingly in the master equation
systems~\cite{Lebowitz,SeifertPRL05,Esposito07,Harris,LiuF2,EspositoPRL10}.
Nevertheless, whether a similar time reversal definition exists
and how to define it are not investigated yet. We must emphasize
that it is not trivial as one might think of at first glance.
Except for positive transition rates, in these master equation
systems there are not quantities such as drift vector and
diffusion matrix that have intuitive rules of transformation as
time reversed~\cite{Risken}. Somewhat surprisingly, here we will
show that an physically relevant answer indeed exists.

The organization of this work is as follows. In sec.~\ref{review},
we briefly review a GIFR in the master equation systems that we
found very recently~\cite{LiuF2}. The reason we use the GIFR
rather than other famous fluctuation relations is its generality.
Additionally, the necessary of extending the conventional time
reversal will be brought forth naturally in deriving the GIFR. In
order to interpret an unknown matrix in this relation, in
sec.~\ref{splitting} we present a generalized time reversal in
these systems as a splitting of rate matrix into reversal and
irreversible parts. The consequences of this new definition will
be discussed in sec.~\ref{discussion}, which includes
reinvestigation of the fluctuation relations from a point of view
of the splitting, generalization of total entropy production and
conditions of the detailed balance. Section~\ref{conclusion} is
the summary.

\section{Review of the GIFR in the master equation
systems} \label{review} Assume a Markovian process with discrete
states and continuous time is described by a master
equation~\cite{LiuF2}
\begin{eqnarray}
\frac{dp_n(t)}{dt}=\left[{\textbf H}(t){\textbf p}(t)\right]_n,
\label{forwardeq}
\end{eqnarray}
where $n$ is the state index which may be a vector, the
$N$-dimensional column vector ${\textbf
p}(t)$$=$$(p_1,\cdots,p_N)^{\rm T}$ is the probability of the
system at individual states at time $t$, and the matrix element
$({\bf H})_{mn}$=$H_{mn}$$>$0 ($m$$\neq$$n$) is the time dependent
or independent rate and  $({\textbf H})_{nn}$=$-$$\sum_{m\neq
n}H_{mn}$. Given a normalized positive column vector ${\textbf
f}(t)$=$(f_1,\cdots,f_N)^{\rm T}$ and a $N$$\times$$N$ matrix
${\textbf A}$ whose elements $({\bf A})_{mn}$$=$$A_{mn}$ ($m\neq
n$) satisfy condition $H_{mn}f_n$$+$$A_{mn}$$>0$  and
$A_{nn}$$=$$-\sum_{m\neq n}A_{mn}$, we found that the inner
product ${\textbf f}^{\rm T}(t'){\textbf v}(t')$ ($t'$$<$$t$) is
$t'$-invariable if the vector $\textbf v(t')=(v_1,\cdots,v_N)^{\rm
T}$  satisfies a perturbed backward equation
\begin{eqnarray}
\frac{dv_n(t')}{dt'}=&-&({\textbf H}^{\rm T}{\textbf
v})_n\nonumber\\&-&f_n^{-1}\left(\partial_{t'}{ \textbf
f}-{\textbf H}{ \textbf f}\right)_nv_n+f_n^{-1} [\left({ \textbf
A} \textbf 1\right)_nv_n-({\textbf A}^{\rm T} {\textbf v})_n],
\label{backwardeq}
\end{eqnarray}
where the final condition $v_n(t)$=$d_n$, and $N$-dimension column
vector ${\textbf 1}=(1,\cdots,1)^{\rm T}$. This is easily proved
by noticing the time differential property and the transpose
property of a matrix. Employing the Feynman-Kac and Girsanov
formulas for discrete jump processes~\cite{LiuF2}, the solution
of~(\ref{backwardeq}) has a path integral representation given by
\begin{eqnarray}
\label{GIFT} \sum_{m=1}^N f_m(0) ^{m,0}\langle e^{-{\int_0^t\cal
J}[\textbf f,\textbf A]({\bf x}(\tau))d\tau }d_{{\textbf
x}(t)}\rangle={\textbf f}^{\rm T}(t) {\textbf d}
\end{eqnarray}
and the integrant in the functional is
\begin{eqnarray}
{\cal J}[{\bf f,A}]&=&f_{\textbf
x(\tau)}^{-1}\left[-\partial_{\tau}\textbf f+\textbf H\textbf
f+\textbf A\textbf 1\right]_{{\textbf x}(\tau)}+{\cal
Q}[f_{\textbf x(\tau)}^{-1}\textbf A]\label{functional}
\end{eqnarray}
with
\begin{eqnarray}
{\cal Q}[\textbf B]&=&-B_{{\textbf x}(\tau){\textbf
x}(\tau)}-\ln\left[1+\frac{B_{{\textbf x}(\tau^+){\textbf
x}(\tau^-)}(\tau)}{H_{{\textbf x}(\tau^+){\textbf
x}(\tau^-)}(\tau)}
\right]\sum_{i=1}^k\delta(\tau-\tau_i),\label{functionalG}
\end{eqnarray}
where $^{m,0}\langle\rm$ $\rangle$ is the expectation over all
trajectories $\textbf x$ generated from the
system~(\ref{forwardeq}) with a fixed state $m$ at initial time 0,
$\textbf x(\tau)$ is the discrete state of the system at instant
time $\tau$, $\textbf x(\tau^{-})$ and $\textbf x(\tau^{+})$
represent the states just before and after a jump occurring at
time $\tau$, respectively, and we have assumed the jumps occur $k$
times for a trajectory. By selecting different ${\bf f}$ and ${\bf
A}$, the GIFR (\ref{GIFT}) may be reduced into different
fluctuation relations in the literature~\cite{LiuF2}.

Although (\ref{backwardeq}) seems complicated, in fact it could be
arranged into a concise form:
\begin{eqnarray}
\frac{dq_{\tilde{n}(s)}}{ds}=[\widetilde{\textbf H}(s)\textbf
q(s)]_{\tilde{n}},\label{timereversal}
\end{eqnarray}
where the elements of the probability ${\bf q}(s)$ are
\begin{eqnarray}
\label{reversedprobdefinition} q_{\tilde{n}}(s)=[{\bf f}(t)^{\rm
T}{\bf d}]^{-1}f_n(t')v_n(t')
\end{eqnarray}
with $s$=$t$$-$$t'$, and the elements of the matrix
$\widetilde{\bf H}(s)$ are
\begin{eqnarray}
\widetilde{H}_{{\tilde n}{\tilde
m}}(s)=f_m^{-1}(t')\left[H_{mn}(t')f_n(t')+A_{mn}(t')\right]
\label{ratetimereversal}
\end{eqnarray}
for $m$$\neq$$n$ and $\widetilde{H}_{{\tilde  m}{\tilde
m}}(s)$$=$$-\sum_{{\tilde  n}\neq{\tilde  m}}\widetilde H_{{\tilde
n}{\tilde  m}}(s)$, and $\tilde n$ represents an index whose
components are the same or the minus of themselves depending on
whether they are even or odd under time reversal $t$$\to$$-t$.
Considering that the parameter $s$ is analogous to a reversed time
and $\widetilde{H}_{{\tilde n}{\tilde m}}(s)$$=$${H}_{nm}(t')$ as
specifically selecting ${\bf A}$ whose elements are the
probability flux $[{\bf J}({\bf
f})]_{mn}$=$(H_{nm}f_m$$-$$H_{mn}f_{n})$, we simply name
(\ref{timereversal}) as generalized time reversal of the original
system (\ref{forwardeq}). However, why ${\bf A}$ is almost
arbitrary instead of equaling probability flux only was not fully
understood previously. Hence, defining the generalized time
reversal from a different point of view seems very essential.

\section{Splitting rate matrix as time reversal}
\label{splitting} Let us begin with two general matrix identities:
\begin{eqnarray}
{[{\textbf M} ({\bf a. \bf b})]}_n-{[{\textbf M}{\bf
a}]}_nb_n+a_n{[{\textbf M}^{\rm
T}{\bf b}]}_n&=&[{\textbf S}^{\rm T}{\bf b}]_n-[{\textbf S}{\textbf 1}]_nb_n,\label{identity1} \\
{[{\textbf M} ({\bf a. b})]}_n+{[{\textbf M}{\bf
a}]}_nb_n-a_n{[{\textbf M}^{\rm T}{\bf b}]}_n&=&[{\textbf J}^{\rm
T}{\bf b}]_n-[{\textbf J}{\bf 1}]_nb_n\label{identity2} ,
\end{eqnarray}
where both ${\bf a}$ and ${\bf b}$ are $N$-dimensional vectors,
${\bf M}$ is $N$$\times$$N$ {\bf matrix with $\sum_{m}
M_{mn}$$=$$0$}, $({\bf a. \bf b})$=$(a_1b_1,\cdots,a_Nb_N)^{\rm
T}$ represents an array multiplication of the vectors, and the
matrixes ${\bf S}$ and ${\bf J}$ are constructed by ${\bf M}$ and
${\bf a}$:
\begin{eqnarray}
({\bf S})_{mn}=M_{nm}a_m+ M_{mn}a_n,\hspace{0.4cm}({\bf
J})_{mn}=M_{nm}a_m-M_{mn}a_n.
\end{eqnarray}
They are symmetric and antisymmetric, respectively. Proving
(\ref{identity1}) and~(\ref{identity2}) is straightforward.

Now we assume that the rate matrix can be splitted into a sum of
``reversible"  and ``irreversible" matrixes, namely ${\bf
H}$=${\bf H}^{\rm rev}$+${\bf H}^{\rm irr}$. Particularly, we
require $H^{\rm irr}_{mn}$$>$$H^{\rm rev}_{mn}$ ($m$$\neq$$n$),
the reason of which will be seen shortly. It is easily to see that
$H^{\rm irr}_{mn}$ is always greater than zero for $H_{mn}$$>$0,
though the sign of $H^{\rm irr}_{mn}$ is indefinite. We further
assume ${\bf H}$ is transformed into ${\bf \widetilde
H}$=${\bf\widetilde H}^{\rm rev}$+${\bf\widetilde H}^{\rm irr}$
under time reversal and
\begin{eqnarray}
\widetilde{H}^{\rm irr}_{\tilde{m}\tilde{n}}(s)={H}^{\rm
irr}_{mn}(t'), \hspace{0.4cm} \widetilde{H}^{\rm
rev}_{\tilde{m}\tilde{n}}(s)=-{H}^{\rm rev}_{mn}(t').
\end{eqnarray}
Obviously, time-reversed matrix ${\bf\widetilde H}$ still
possesses rate interpretation. Using these definitions and
identities we rewrite the right-hand side of~(\ref{timereversal})
as
\begin{eqnarray}
\label{expandingreversedH} [\widetilde{\bf H}{\bf
q}]_{\tilde{n}}&=&\frac{1}{{\bf f}(t){\bf d}^{\rm
T}}\{[\widetilde{\bf H}^{\rm irr}({\bf
f.v})]_{\tilde{n}}+[\widetilde{\bf H}^{\rm rev}({\bf
f.v})]_{\tilde{n}}\}\\
&=&\frac{1}{{\bf f}(t){\bf d}^{\rm T}}\{({\textbf H}^{\rm
T}{\textbf v})_n-({\textbf H}{\textbf
f})_nv_n-[(\textbf J^{\rm irr}-\textbf S^{\rm rev})\textbf 1]_nv_n\nonumber\\
&&\hspace{1.2cm}+[(\textbf J^{\rm irr}-\textbf S^{\rm rev})^{\rm
T}\textbf v]_n\},\nonumber
\end{eqnarray}
where
\begin{eqnarray}
\label{partialfluxdef}
({\textbf S}^{\rm rev})_{mn}&=& H^{\rm rev}_{nm}f_m+H^{\rm rev}_{mn}f_n,\\
({\textbf J}^{\rm irr})_{mn}&=& H^{\rm irr}_{nm}f_m-H^{\rm
irr}_{mn}f_n.
\end{eqnarray}
Comparing (\ref{expandingreversedH}) with (\ref{backwardeq}), we
immediately find that
\begin{eqnarray}
\label{Amatrixdef1} [{\bf A(f)}]_{mn}&=&({\bf J}^{\rm
irr})_{mn}-({\bf S}^{\rm
rev})_{mn}\\
\label{Amatrixdef2}&=&[{\bf J(f)}]_{mn}-2H^{\rm rev}_{nm}f_m,
\end{eqnarray}
(\ref{Amatrixdef2}) implies why the matrix ${\bf A}$ is almost
arbitrary: the reversible matrix ${\bf H}^{\rm rev}$ to be
determined is responsible for this freedom. In addition, we also
notice that the condition of $H_{mn}f_n$$+$$A_{mn}$$>$0 mentioned
in last section is identical with $H^{\rm irr}_{mn}$$>$$H^{\rm
rev}_{mn}$ here.

In practice, we may prior know the matrix ${\bf A}$ or
time-reversed rate matrix $\widetilde {\bf H}$. Under these
circumstances, the splitting can be constructed conversely as
\begin{eqnarray}
\label{splitdefbyArev} H^{\rm
rev}_{nm}=(H_{nm}f_m-H_{mn}f_n-A_{mn})/2f_m,\\
\label{splitdefbyAirr}H^{\rm
irr}_{nm}=(H_{nm}f_m+H_{mn}f_n+A_{mn})/2f_m,
\end{eqnarray}
or
\begin{eqnarray}
\label{splitdefbyreversedHrev} H^{\rm
rev}_{mn}(t')=[H_{mn}(t')-\widetilde
H_{\tilde{m}\tilde{n}}(s)]/2,\\
\label{splitdefbyreveredHirr} H^{\rm
irr}_{mn}(t')=[H_{mn}(t')+\widetilde H_{\tilde{m}\tilde{n}}(s)]/2.
\end{eqnarray}
Regardless of how a splitting of the rate matrix is achieved,
substituting (\ref{Amatrixdef1}) into the integrand
(\ref{functional}) we can obtain its new expression given
by~\footnote{In the following we will alternatively use $\cal
J[{\bf f},{\bf A}]$ and $\cal J[{\bf f},{\bf H}^{\rm irr},{\bf
H}^{\rm rev}]$ without explicit statement, which would not result
in confusion because of (\ref{Amatrixdef1}).}
\begin{eqnarray}
\label{newfunctional1} {\cal J}[{\bf f},\textbf H^{\rm
irr},\textbf H^{\rm rev}] &=&-\frac{d}{d\tau}\ln f_{{\bf
x}(\tau)}(\tau)+{\cal J}[\textbf 1,\textbf H^{\rm irr},\textbf
H^{\rm rev}],
\end{eqnarray}
where the second term on the right-hand side is
\begin{eqnarray}
\label{trajectorymediaentropy} -2\sum_{m\neq {\bf x}(\tau)} H^{\rm
rev}_{m{\bf x}(\tau)} \label{newfunctional2}-\ln[\frac{H^{\rm
irr}_{{\bf x}(\tau^-){\bf x}(\tau^+)}-H^{\rm rev}_{{\bf
x}(\tau^-){\bf x}(\tau^+)}}{H^{\rm irr}_{{\bf x}(\tau^+){\bf
x}(\tau^-)}+H^{\rm rev}_{{\bf x}(\tau^+){\bf
x}(\tau^-)}}]\sum^k_{i=1}\delta(\tau-\tau_i).
\end{eqnarray}
(\ref{newfunctional1}) is a consequence of the arbitrariness of
${\bf f}$ in~(\ref{reversedprobdefinition}), a brief explanation
of which is given in Appendix A. Noting
(\ref{trajectorymediaentropy}) also satisfies a fluctuation
relation if the initial distribution of the system is uniform.
(\ref{newfunctional1}) and (\ref{trajectorymediaentropy}) are the
central results of this work.

\section{Discussion}
\label{discussion} Because the physical interpretation of the
GIFR~(\ref{GIFT}), its connection with the other fluctuation
relations, and its detailed version have been partially
investigated~\cite{LiuF2}, in the remainder we only present
several new results obtained from a point of view of the
splitting.

\subsection{Total entropy production rate}
If ${\bf f}$ is the system's probability vector ${\bf p}$ and the
splitting is prior known from physical consideration, e.g., time
reversibility below, (\ref{newfunctional1}) is the balance
equation of trajectory entropy: the first term on the right-hand
side is the change in trajectory system entropy, the second term
is the change in trajectory environmental entropy along a
specified trajectory, and the left-hand side is the total
trajectory entropy. This interpretation has been widely
accepted~\cite{SeifertPRL05,Esposito07,Harris,EspositoPRL10,GeQian}.
To our knowledge, however, the expression of the trajectory
environmental entropy (\ref{newfunctional2}) that involves both
even and odd discrete variables is firstly given here. This new
formula also reminds us that the total entropy production for the
master equation systems with irreversible and reversible rate
matrixes is
\begin{eqnarray}
\label{entropyprod} \langle{\cal J}\rangle =
&-&2\sum_{m\neq n}\sum_nH^{\rm rev}_{nm}p_m\nonumber\\
&-&\sum_{m\neq n}\sum_n(H^{\rm irr}_{nm}+H^{\rm
rev}_{nm})p_m\ln\frac{(H^{\rm irr}_{mn}-H^{\rm
rev}_{mn})p_n}{(H^{\rm irr}_{nm}+H^{\rm rev}_{nm})p_m}\geq 0.
\end{eqnarray}
The inequality may be proved easily by using ($x$$-1$)$\ge$$\ln x$
or using the Jensen inequality for the GIFR~(\ref{GIFT}). The
expression of (\ref{entropyprod}) is significantly distinct from
the classical entropy production formula given by
Schnakenberg~\cite{Schnakenberg} that involves only even
variables.

\subsection{Conditions of the detailed balance}
For a time-independent master equation system that has equilibrium
state ${\bf p}^{\rm eq}$, we may physically require it invariable
under time-reversal, namely, $\widetilde{\bf H}$$=$${\bf H}$.
According to~(\ref{splitdefbyreversedHrev})
and~(\ref{splitdefbyreveredHirr}) an unique splitting is then
\begin{eqnarray}
\label{splitdefbyreversibleH} H^{\rm
rev}_{mn}=(H_{mn}-H_{\tilde{m}\tilde{n}})/2,\hspace{0.4cm} H^{\rm
irr}_{mn}=(H_{mn}+ H_{\tilde{m}\tilde{n}})/2.
\end{eqnarray}
Under this circumstance the irreversible and reversible parts have
distinctive properties:
\begin{eqnarray}
\label{reversible} H^{\rm rev}_{nm}=-H^{\rm rev}_{{\tilde
n}{\tilde m}},\hspace{0.4cm} H^{\rm irr}_{nm}=H^{\rm irr}_{{\tilde
n}{\tilde m}}.
\end{eqnarray}
Obviously, if the discrete master equation system involves only
even states, which is exclusively the object investigated in
numerous references~\cite{Gardiner,Risken,Oppenheim,Kampen}, the
reversible rate matrix must vanish. Because the entropy production
is zero when the system is in equilibrium state, according to
(\ref{entropyprod}), all terms on its right hand side must vanish.
Hence we obtain conditions of detailed balance on the rate matrix
and the state which are respectively
\begin{eqnarray}
\label{detailbalancerev} ({\textbf S}^{\rm rev})_{mn}&=&H^{\rm
rev}_{nm}p^{\rm eq}_m+ H^{\rm rev}_{mn}p^{\rm eq}_n=0,\\
\label{detailbalanceirr} ({\textbf J}^{\rm irr})_{mn}&=&H^{\rm
irr}_{nm}p^{\rm eq}_m-H^{\rm irr}_{mn}p^{\rm eq}_n=0,\\
\label{steadycondition}\sum_{m\neq n} H^{\rm rev}_{mn}&=&0.
\end{eqnarray}
We must emphasize that, except for the last equation
(\ref{steadycondition}) that is from the steady-state requirement
and was ignored in previous
literature~\cite{Gardiner,Risken,Oppenheim,Kampen}, the former two
equations are fully equivalent with the conventional detailed
balance condition
\begin{eqnarray}
\label{classicdetailbalance}
 H_{{\tilde n}{\tilde m}}p^{\rm
eq}_m=H_{mn}p^{\rm eq}_n,
\end{eqnarray}
which may be easily checked using (\ref{splitdefbyreversibleH}).
We see an unusual side of a time-reversible master equation system
with nonvanishing $\textbf H^{\rm rev}$: the probability flux
$({\bf J})_{mn}$ between two states $m$ and $n$ may be not zero
even if the system is in an equilibrium state; both the
irreversible and reversible rate matrixes have contributes to the
total entropy production. The latter point can be seen more
clearly when we rewrite (\ref{entropyprod}) as
\begin{eqnarray} \label{entropyproddetailed}
\langle{\cal J}\rangle =&-&\sum_{m\neq n}\sum_nH^{\rm irr}_{nm}p_m\ln\frac{H^{\rm irr}_{mn}p_n}{H^{\rm irr}_{nm}p_m}\\
&-&\sum_{m\neq n}\sum_nH^{\rm rev}_{nm}p_m\ln(-\frac{H^{\rm
rev}_{mn}p_n}{H^{\rm rev}_{nm}p_m}).\nonumber
\end{eqnarray}
by taking (\ref{detailbalancerev})-(\ref{steadycondition}) into
account. To our best knowledge, there are very fewer stochastic
jump processes with both even and odd variables in the literature.
In Appendix B we present a simple mathematical model to exemplify
their intriguing features.

\subsection{Jarzynski and Hatano-Sasa equalities}
For a time-dependent master equation system, we may select the
simplest case with ${\bf A}$=0. Then~(\ref{splitdefbyArev})
and~(\ref{splitdefbyAirr}) become
\begin{eqnarray}
\label{splittingwith0A} {\cal{H}}^{\rm
rev}_{nm}=\frac{1}{2f_m}(H_{nm}f_m-H_{mn}f_n),\\ {\cal{H}}^{\rm
irr}_{nm}=\frac{1}{2f_m}(H_{nm}f_m+H_{mn}f_n).
\end{eqnarray}
We used a new notation ${\cal H}$ instead of $H$ to indicate the
speciality of this splitting. Under this case,
(\ref{newfunctional1}) is
\begin{eqnarray}
\label{newfunctional1neqss} {\cal J}[{\bf f},{\bf A}=0]
&=&-\frac{d}{d\tau}\ln f_{{\bf x}(\tau)}(\tau)+{\cal J}[\textbf
1,{\cal H}^{\rm irr},{\cal H}^{\rm rev}],
\end{eqnarray}
and
\begin{eqnarray} \label{JEHEfunctional}  {\cal J}[{\bf
1},{\cal H}^{\rm irr}, {\cal H}^{\rm rev}]=&&-\frac{1}{f_{{\bf
x}(\tau)}}\sum_{m\neq {{\bf x}}(\tau)}[H_{{\bf
x}(\tau)m}f_m-H_{m{\bf x}(\tau)}f_{{\bf
x}(\tau)}]\nonumber\\
&&-\ln\frac{ f_{\bf x(\tau^-)}}{f_{\bf x(\tau^+)}
}\sum^k_{i=1}\delta(\tau-\tau_i).
\end{eqnarray}
If one further selects ${\bf f}$ to be the instant equilibrium
solution ${\bf p}^{\rm eq}(t)$ or the instant nonequilibrium
steady-state solution ${\bf p}^{\rm ss}(t)$ of the master equation
if it has, the GIFR is respectively the famous Jarzynski equality
about dissipated work~\cite{JarzynskiPRL97,JarzynskiPRE97} or
Hatano-Sasa equality about excess entropy~\cite{HatanoSasa}.
Noting the first term of the right-hand side of the above equation
in both cases vanishes because of the steady-state condition. It
is worth pointing out that for the former equality, the detailed
balance conditions~(\ref{detailbalancerev})
and~(\ref{detailbalanceirr}) imply that such a splitting is
trivial because of ${\cal{H}}^{\rm rev}_{nm}(t)$=$H^{\rm
rev}_{nm}(t)$ and ${\cal{H}}^{\rm irr}_{nm}(t)$=$H^{\rm
irr}_{nm}(t)$, and (\ref{JEHEfunctional}) has other three
different but equivalent expressions:
\begin{eqnarray}
\label{JEHEfunctionalirreversible} \ln \frac{ p^{\rm eq}_{{\bf
x(\tau^-)} }}{p^{\rm eq}_{{\bf x(\tau^+)} }}&&=\ln \frac{ H^{\rm
irr}_{{\bf x(\tau^-)}{\bf x(\tau^+)}}}{H^{\rm irr}_{{\bf
x(\tau^+)}{\bf x(\tau^-)}}}=\ln [-\frac{ H^{\rm rev}_{{\bf
x(\tau^-)}{\bf x(\tau^+)}}}{H^{\rm rev}_{{\bf x(\tau^+)}{\bf
x(\tau^-)}}}]\nonumber\\&&=\ln \frac{ H^{\rm irr}_{{\bf
x(\tau^-)}{\bf x(\tau^+)}}-H^{\rm rev}_{{\bf x(\tau^-)}{\bf
x(\tau^+)}}}{H^{\rm irr}_{{\bf x(\tau^+)}{\bf x(\tau^-)}}+H^{\rm
rev}_{{\bf x(\tau^+)}{\bf x(\tau^-)}}}.
\end{eqnarray}

On the other hand, if $\bf A1$$=$$0$ the
integrand~(\ref{functional}) can be decomposed into
\begin{eqnarray}
\label{decompositionJ} {\cal J}[{\bf f}, {\bf A}]={\cal J}[{\bf
f},{\bf 0}]+{\cal Q}[f_{\textbf x(\tau)}^{-1}\textbf A].
\end{eqnarray}
This relationship is very intriguing. In addition that the three
terms above satisfy integral fluctuation relation~\cite{LiuF2}
simultaneously, all of them have important physical implications
in the nonequilibrium steady-state
thermodynamics~\cite{HatanoSasa,Speck,EspositoPRL10,GeQian,Oono}.
Here we do not repeat previous interpretations but present a
simple understanding from a point of view of the splitting. For a
nonequilibrium master equation system with ${\bf H}(t)$$=$${\bf
H}^{\rm irr}(t)$ and assuming it has unique steady-state as
external time-dependent parameter fixed, we have
\begin{eqnarray}
\label{entropyproductionfaces1} {\cal J}[{\bf p},{\bf J(p)}
]=&-&\frac{d}{d\tau}\ln \frac{p_{{\bf x}(\tau)}}{{p}_{{\bf
x}(\tau)}^{\rm ss}} +{\cal J}[{\bf p}^{ss},{\bf J(p^{\rm ss})}]\\
&=&-\frac{d}{d\tau}\ln\frac{ p_{{\bf x}(\tau)}}{{p}_{{\bf
x}(\tau)}^{\rm ss}}+{\cal J}[{\bf p}^{ss},{\bf 0}]+{\cal
Q}[\frac{{\bf J}(p^{\rm
ss})}{p_{\bf x(\tau)}^{\rm ss}}]\label{entropyproductionfaces2}\\
&=&-\frac{d}{d\tau}\ln p_{{\bf x}(\tau)}+{\cal J}[{\bf 1},{\cal
H}^{\rm irr}, {\cal H}^{\rm rev}]+{\cal Q}[\frac{{\bf J}(p^{\rm
ss})}{p_{\bf x(\tau)}^{\rm ss}}],\label{entropyproductionfaces3}
\end{eqnarray}
where (\ref{newfunctional1}), (\ref{newfunctional1neqss}),
(\ref{decompositionJ}) and steady-state condition ${\bf J(p^{\rm
ss})1}$=0 are used. (\ref{entropyproductionfaces2}) implies that
trajectory total entropy is the sum of trajectory relative
entropy, trajectory excess entropy~\cite{HatanoSasa}, and
trajectory housekeeping entropy~\cite{Speck}. Particularly, the
sum of the first two terms in (\ref{entropyproductionfaces3}) that
was called as nonadiabatic trajectory entropy~\cite{EspositoPRL10}
is just ${\cal J}[{\bf p},{\cal H}^{\rm irr}, {\cal H}^{\rm
rev}]$, the ensemble average of which is
\begin{eqnarray}
\langle{\cal J}[{\bf p},{\cal H}^{\rm irr}, {\cal H}^{\rm
rev}]\rangle = &-&\sum_{m\neq n}\sum_nH_{nm}p_m\ln\frac{p^{\rm
ss}_mp_n}{p_n^{\rm ss}p_m}\geq 0
\end{eqnarray}
according to (\ref{entropyprod}). In Ref.~\cite{GeQian} this
quantity was also named as free energy dissipation.

\section{Conclusion}
\label{conclusion} Motivated by the important idea of defining
time reversal as a splitting of drift vector in continuous
diffusion process, we present the same effort in the master
equation system with discrete states. Different from the former,
we define the time reversal in this system as a splitting of the
rate matrix into irreversible and reversible parts. Even that we
find very analogous formulas and results are revealed in these two
systems. e.g. (\ref{trajectorymediaentropy}) corresponding (7.6)
in the work of Chetrite and Gawedzki~\cite{Chetrite} or (28) in
the work of us~\cite{LiuF3}. The advantages of introducing this
definition are obvious. First, we explain the origin of the matrix
$\bf A$ in the GIFR, which was somewhat mysterious to us before
starting this work. Second, a variety of fluctuation relations in
the master equation systems are unified into various rate matrix
splitting. This point was not acknowledged previously.
Additionally, the relationships among these fluctuation relations
become very clear from this splitting viewpoint,
e.g.~(\ref{entropyproductionfaces2})
and~(\ref{entropyproductionfaces3}). Finally, this definition also
reminds us the importance of the reversible part of the rate
matrix. For instance, the expression of the total entropy
production must be modified appropriately. To our knowledge, its
existence and implications in the master equation systems were
almost completely ignored for a very long time.\\

This work was supported in part by State Key Laboratory of
Software Development Environment SKLSDE-2011ZX-20 and the National
Science Foundation of China under Grant No. 11174025.

\appendix
\setcounter{section}{1}
\section*{Appendix A: Derivation of (\ref{newfunctional1}) and (\ref{newfunctional2})}
For simplicity in notations, we study a specific case of
(\ref{backwardeq}) with a final condition is $d_n$=1. Under this
circumstance the perturbed backward master equation has the
time-reversal explanation (\ref{timereversal}) and
\begin{eqnarray}
\label{reversedprobdefinition1} q_{\tilde n}(s)=f_n(t')v_n(t').
\end{eqnarray}
Noting that the initial condition of $q_{\tilde n}$ is now
$f_n(t)$. It is worth emphasizing that we can replace the vector
$\bf f$ above by other normalized positive vectors, e.g. $\bf
c$=$(1/N,\cdots,1/N)$. Then we have a new relation
\begin{eqnarray}
\label{reversedprobdefinition2} q_{\tilde n}(s)=c_nu_n(t'),
\end{eqnarray}
where $u_n$ satisfies the same (\ref{backwardeq}) except that
$f_n$ therein including those in the matrix ${\bf A(f)}$
(\ref{Amatrixdef1}) are substituted by $c_n$ and the finial
condition becomes $u_n(t)$=$f_n(t)/c_n$. According to
(\ref{reversedprobdefinition1}), (\ref{reversedprobdefinition2}),
and the path integral representation (\ref{GIFT}), we immediately
obtain
\begin{eqnarray}
{ }^{n,0}\langle e^{-{\int_0^t\cal J}[\textbf c,\textbf A(\textbf
c)]({\bf x}(\tau))d\tau }\frac{f_{{\textbf x}(t)}(t)}{f_n(0)}
\rangle= ^{n,0}\langle e^{-{\int_0^t\cal J}[\textbf f,\textbf
A]({\bf x}(\tau))d\tau } \rangle
\end{eqnarray}
Because of ${\cal J}[\textbf c,\textbf A(\textbf c)]$=${\cal
J}[\textbf 1,\textbf A(\textbf 1)]$ (\ref{newfunctional1}) is
proved.

\section*{Appendix B: A discrete jump model with both even and odd variables }
\begin{figure}
\includegraphics{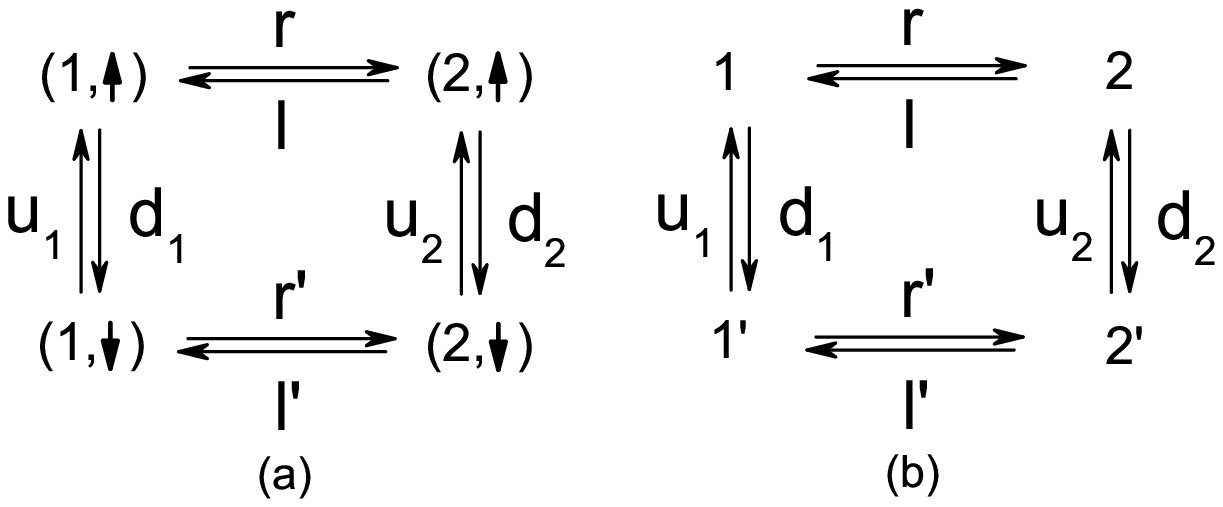}
\caption{(a) A discrete jump process with both even (1 and 2) and
odd ($\uparrow$ and $\downarrow$) variables, (b) jump process with
only even variables (1, 1$'$, 2, and 2$'$). The letters
$r,l,r',l',u_1,d_1,u_2,d_2$ ($>$0) are respective rates.
}\label{figure1}
\end{figure}
To show the unusual features of jump processes with both even and
odd variables, we develop a simple model; see
Fig.~(\ref{figure1})(a). Assuming that the variable $i$=1 and $2$
are even and $\uparrow$ and $\downarrow$ are odd,  under time
reversal ($t$$\rightarrow$$-t$) we have
$\widetilde{(i,\uparrow)}=(i,\downarrow)$ and
$\widetilde{(i,\downarrow)}=(i,\uparrow)$. Although this model has
eight rate parameters, they would not to be independent of each
other if we required the model to have a steady-state solution
satisfying the detailed balance principle. Using
(\ref{detailbalancerev})-(\ref{steadycondition}) and performing a
simple calculation, we find these rates must obey the following
three constraints:
\begin{eqnarray}
\label{contraintsevenodd1}
l r-l' r'=0,\\
r-r'+d_1-u_1=0,\label{contraintsevenodd2}\\
l-l'+d_2-u_2=0.\label{contraintsevenodd3}
\end{eqnarray}
The first equation is a consequence of (\ref{detailbalancerev}) or
(\ref{detailbalanceirr}), while the last two equations are from
(\ref{steadycondition}). The equilibrium solutions can be easily
obtained, which are
\begin{eqnarray}
\label{equilibriumsolutionsevenodd1}
p^{\rm eq}_{(1,\uparrow)}=p^{\rm eq}_{(1,\downarrow)}=(l+l')/2(l+l'+r+r'),\\
\label{equilibriumsolutionsevenodd2}p^{\rm
eq}_{(2,\uparrow)}=p^{\rm
eq}_{(2,\downarrow)}=(r+r')/2(l+l'+r+r').
\end{eqnarray}
In addition, the total entropy
production~(\ref{entropyproddetailed}) is
\begin{eqnarray}
\langle {\cal J}\rangle
&=&[d_1p_{(1,\uparrow)}-u_1p_{(1,\downarrow)}]\ln\frac{p_{(1,\uparrow)}}{p_{(1,\downarrow)}}+[d_2p_{2,\uparrow)}-u_2p_{(2,\downarrow)}]\ln\frac{p_{(2,\uparrow)}}{p_{(2,\downarrow)}}\nonumber\\
&+&[rp_{(1,\uparrow)}-lp_{(2,\uparrow)}]
\ln\frac{(r+r')p_{(1,\uparrow)}}{(l+l')p_{(2,\uparrow)}}+[r'p_{(1,\downarrow)}-l'p_{(2,\downarrow)}]
\ln\frac{(r+r')p_{(1,\downarrow)}}{(l+l')p_{(2,\downarrow)}},
\end{eqnarray}
where $p_{(i,\uparrow)}$ and $p_{(i,\downarrow)}$ are the
transient probabilities of the model at distinct states started
from an initial distribution. Reader may easily check that, if
$p$=$p^{\rm eq}$, the probability currents ${\bf J}$ or the terms
before these logarithmic functions above are nonzero.

\begin{figure}
\includegraphics{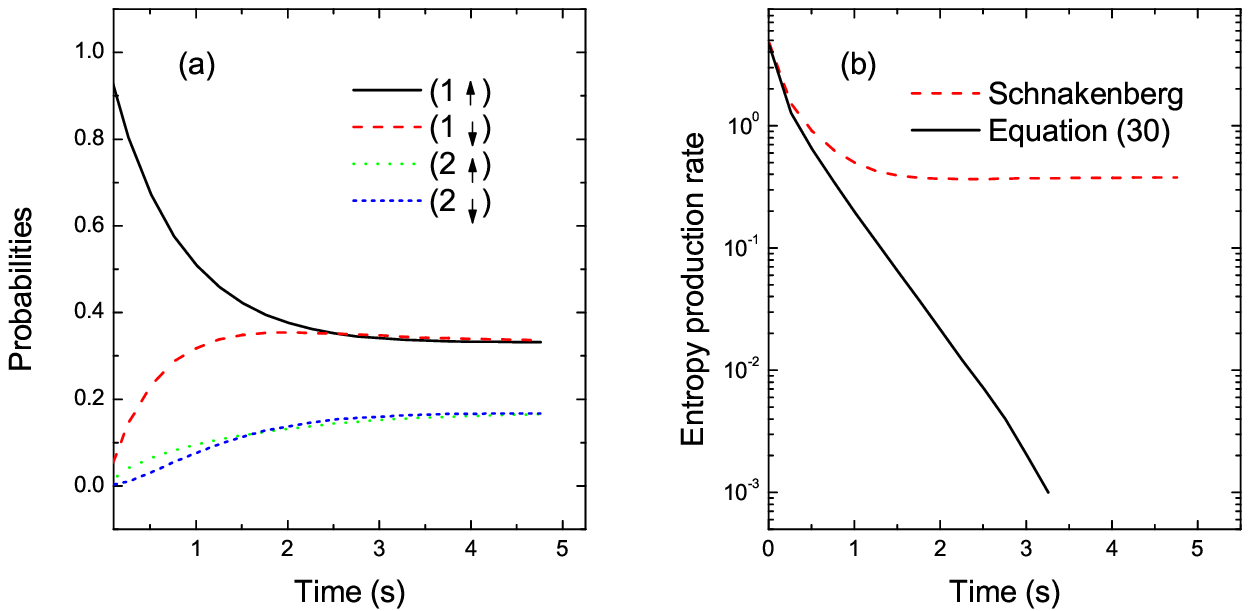}
\caption{(a) Time evolution of the probabilities at the four
states in Fig.~(\ref{figure1})(a) and (b) the entropy production
rates calculated by the classical Schnakenbergs
formula~(\ref{SchnakenbergEP}) and (\ref{entropyproddetailed})
proposed in this work. We choose $l$=1s$^{-1}$, $r$=0.2s$^{-1}$,
$l_1$=0.4s$^{-1}$, $r_1$=0.5s$^{-1}$; $d_1$=0.7s$^{-1}$;
$u_1$=0.4s$^{-1}$, $d_2$=0.2s$^{-1}$, and $u_2$=0.8s$^{-1}$, which
satisfy (\ref{contraintsevenodd1})-(\ref{contraintsevenodd3}). The
initial conditions are $p_{(1,\uparrow)}$=1.0, and the others
vanish. }\label{figure2}
\end{figure}

It would be interesting to compare these results with those
obtained in a conventional jump processes with only even
variables, e.g. Fig.~(\ref{figure1})(b). Its states are
respectively marked by the variables 1, 2, 1$'$, and 2$'$, all of
which are even under time reversal. Because of $p^{\rm
eq}(i,\uparrow)$=$p^{\rm eq}(i,\downarrow)$ in the model (a), here
we additionally require $p^{\rm eq}(i)$=$p^{\rm eq}(i')$. Hence,
we obtain three constraints on the rates given by
\begin{eqnarray}
d_1-u_1=0,\\
d_2-u_2=0,\\
l r'-l' r=0,
\end{eqnarray}
the equilibrium solutions
\begin{eqnarray}
p^{\rm eq}_1=p^{\rm eq}_{1'}=l/2(r+l),\\
p^{\rm eq}_2=p^{\rm eq}_{2'}=r/2(r+l),
\end{eqnarray}
and the classical total entropy production~\cite{Schnakenberg}
\begin{eqnarray}
\label{SchnakenbergEP} \langle {\cal J}\rangle_c
&=&(d_1p_1-u_1p_{1'})\ln(\frac{d_1p_{1}}{u_1p_{1'}})+(d_2p_{2}-u_2p_{2'})\ln(\frac{d_2p_{2}}{u_2p_{2'}})\nonumber\\
&+&(rp_{1}-lp_{2} ) \ln(\frac{rp_{1}}{lp_{2}})
+(r'p_{1'}-l'p_{2'}) \ln(\frac{r'p_{1'}}{l'p_{2'}}),
\end{eqnarray}
where $p_{i}$ and $p_{i'}$ ($i=$1,2) are the transient
probabilities of the model (b) at individual states started from
an initial distribution. Obviously, in the model (a) if one did
not take the odd variable into account and naively used
(\ref{SchnakenbergEP}), namely replacing $p_i$ and $p_{i'}$
therein by $p_{(i,\uparrow)}$ and $p_{(i,\downarrow)}$,
respectively, he or she would find that the classical formula
would not vanish as the system reaches the equilibrium states
(\ref{equilibriumsolutionsevenodd1}) and
(\ref{equilibriumsolutionsevenodd2}). A numerical example to
confirm these results is shown in Fig.~(\ref{figure2}).

\end{document}